\documentclass[journal=jacsat,manuscript=article]{achemso}
\usepackage{subcaption}
\usepackage{caption}
\usepackage{soul,color}
\usepackage{graphicx}
\usepackage{bm}

\newcommand{\bea}{\begin{eqnarray}}
\newcommand{\eea}{\end{eqnarray}}
\usepackage[colorlinks=true, citecolor=blue, linkcolor=blue, urlcolor=blue]{hyperref}
\usepackage[utf8]{inputenc}
\usepackage[T1]{fontenc}
\usepackage{mathptmx}
\usepackage{chemformula} 
\usepackage{comment}
\author{Dibyendu Sardar}
\affiliation{JILA, University of Colorado, Boulder, Colorado 80309, USA}
\author{John L. Bohn} \email{bohn@murphy.colorado.edu}
\affiliation{JILA, University of Colorado, Boulder, Colorado 80309, USA}

\title[]{Sticking lifetime of ultracold CaF molecules in triplet interactions  }

\date{\today}
 \begin{document}

\begin{abstract}
A six-dimensional potential energy surface is constructed for the spin-polarized triplet state of CaF-CaF by  \textit{ab initio}  calculations at the CCSD(T) level of theory, followed by Gaussian process interpolation. The potential is utilized to calculate the density of states for this bi alkaline-earth-halogen system where we find the value  0.038 $\mu$K$^{-1}$, implying a mean resonance spacing of 26 $\mu$K in the collision complex.  This value implies an associated Rice-Ramsperger-Kassel-Marcus lifetime of 18 $\mu$s, thus predicting long-lived complexes in collisions at ultracold temperatures. 
\end{abstract}

\maketitle

\section{Introduction}
The encounter between two CaF molecules, both in their $^2\Sigma$ ground state, can proceed through either singlet or triplet interactions of the four-body system Ca$_2$F$_2$.  In a previous paper \cite{sardar2022}, we developed the singlet scattering surface for this system, identifying the reaction 2CaF $\rightarrow$ CaF$_2$ + Ca as barrierless and exothermic by   $\sim 4300$ cm$^{-1}$.  Thus CaF molecules that are not spin-polarized may be expected to encounter one another on this surface and quickly undergo chemical reactions.  

By contrast, spin-polarized CaF molecules will initially meet only on the triplet surface, which can not lead directly  to chemical reactions in low-temperature gases.  In particular, ultracold gases of spin-polarized CaF molecules \cite{cheuk2020observation} might naively be expected to be completely immune to chemical reaction.  The triplet surface is therefore an important starting point for understanding ultracold reactions of this radical, and we construct such a surface in this article.

In practice, spin-polarized CaF molecules will not simply scatter elastically in a simple way.  For one thing,  the triplet surface is coupled to the singlet surface via interactions such as spin-orbit coupling and spin-rotation coupling.  Also significantly at ultracold temperatures, diatomic molecules are often observed to vanish into long-lived complex states, a phenomenon known as ``sticking'' \cite{Bause23_JPCA}.  Long dwell times in this complex can not only appear as a loss mechanism, they can also enhance even weak couplings to the singlet surface, thereby further reducing the immunity of spin-polarized CaF molecules to reaction.  In this paper we therefore use the triplet surface to estimate this lifetime in the RRKM approximation, finding it to be 18 $\mu$s.

\section{\textit {ab initio} method of calculation}
The \textit{ab initio} electronic structure calculations are performed using the MOLPRO 2012.1 software package \cite{werner2008}. The calculations start by first computing spin-restricted Hartree-Fock (HF) wave functions. Thereafter, the RHF wave functions are used as an initial guess for the next coupled-cluster calculations including singles, doubles, and perturbative triples excitation [CCSD(T)] calculation. This approach is favorable for the short range physics considered, can be described by a single determinant, and is better applicable to the non-reactive triplet surface of CaF-CaF than the MRCI method.

The ground state of a Ca atom possesses two valance electrons in the 4s atomic orbitals and eight sub-valance electrons in the 3s3p orbitals.  We correlate only the outer core electrons 4s of Ca, and therefore, we consider a pseudopotential based correlation-consistent polarized weighted core valence triple-$\zeta$ basis set cc-pw-CVTZ-PP \cite{kirk2017}. The inner core electrons are replaced  by an effective core potential ([ECP10MDF]) \cite{lim2005} of the Stuttgart / Koeln group. By contrast, we use the basis set for the atom F = aug-cc-PVTZ \cite{kendall1992} which is a correlation-consistent valance triple zeta basis and belongs in the augmented class to treat 2s2p electrons explicitly where 1s electrons of F remains uncorrelated. We do not correct for the basis set superposition error and the justification is given in the next following subsection\ref{bsie}.
\begin{figure}[htp]
    \centering
    \includegraphics[width=9cm]{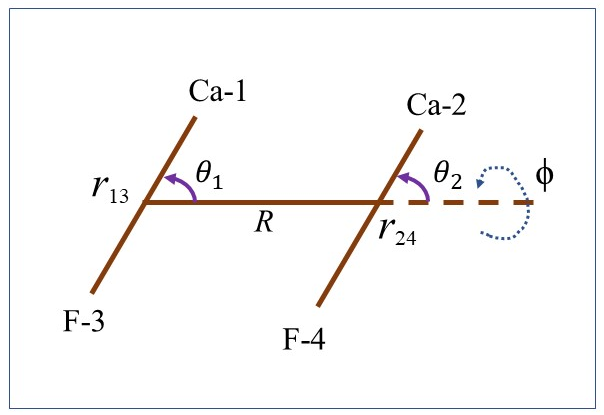}
    \caption{Asymptotic arrangement of the triplet CaF-CaF in the Jacobi coordinates. The numbers 1-4 represent the labeling of the concerned atoms Ca and F. }
    \label{fig1}
\end{figure}
\subsection{Diatomic species}

First,  we study the relevant diatomic species  CaF, Ca$_2$, and F$_2$ to verify our current method of calculation and the used basis sets. Table.\ref{tb1} compares the diatomic properties in terms of the equilibrium bond length ($r_e$) and depth of the well $D_e$ to previously reported theoretical and experimental results. In this table, we show the results of our RHF-CCSD(T) calculations. All the bond lengths are expressed in Bohr and the well depths are in cm$^{-1}$.  The experimental uncertainties are given for those where the values are reported. From Table.\ref{tb1} it is clear that the equilibrium bond length for each of the diatomic systems is in agreement with the literature value with an uncertainty less than 1\% and we accept this as adequate. However, our method underestimates the depth of the well by $\sim 500$ cm$^{-1}$ for both CaF and F$_2$ molecular systems. Therefore, we consider this to estimate  the uncertainty in our current method of calculations. The adequacy of the basis set is discussed in the context of the four-body surface, below.

\begin{table}[h!]
\centering
\caption{Optimized molecular parameters of CaF, F$_2$, and Ca$_2$ where equilibrium bond lengths ($r_e$) are in Bohr and depth of the well ($D_e$) in cm$^{-1}$. }
\begin{tabular}{ |c|c|c|c| } 

\hline
 System & Symmetry & $r_e$ & $D_e$ \\
\hline
CaF &$^2\Sigma$ & 3.70  & 43614  \\ 
theory\cite{hou2018}&, &3.69 &  44203  \\ 
Expt\cite{book}&, &  3.71 & -  \\ 
\hline
F$_2$ &$^1\Sigma$ & 2.68  & 12738 \\ 
theory\cite{meyer2011chemical}&, &2.66 &  12880  \\ 
Expt\cite{chen1985negative}&, &  2.67 & 13410  \\ 
\hline
Ca$_2$ &$^1\Sigma$ & 8.20  & 1097 \\ 
theory\cite{allard2003}&, &8.08 &  1102  \\ 
Expt\cite{balfour1975}&, &  8.09 & 1075 $\pm$ 150  \\ 
\hline
\end{tabular}
\label{tb1}
\end{table}

\subsection{Energies of the possible asymptotes}
\begin{table}[h!]
\centering
\caption{The energy of the four possible asymptotic arrangements with the associated symmetries is considered. The zero of the energy is considered at CaF-CaF asymptote. }
\begin{tabular}{ |c|c|c| } 
\hline
asymptotic arrangements & symmetry & energy (cm$^{-1}$) \\
 \hline
 CaF + CaF & $^2\Sigma + ^2\Sigma$ & 0 \\ 
 \hline
 Ca$_2$ + F$_2$ & $^1\Sigma + ^1\Sigma$  & +73653 \\ 
 \hline
 Ca$_2$F + F & $^2B + ^2P$ & +34483 \\ 
 \hline
CaF$_2$ + Ca & $^1A^{'} + ^1S$ & -4922 \\  
\hline
CaF$_2$ + Ca &$^3A^{''} + ^1S$& +10313 \\ 
\hline
\end{tabular}
\label{tb_2}
\end{table}
Here we discuss the energies of the various dissociation possibilities with respect to the asymptotic energy of triplet CaF-CaF. 
There are four possible asymptotes that need to be considered in the domain of cold collisions between two calcium monofluoride molecules on the triplet surface. These asymptote are: Ca$_2$ + F$_2$, Ca$_2$F + F, CaF$_2$ + Ca, and CaF + CaF. The energies of these possible asymptotes are tabulated in Table.\ref{tb_2} where the zero of the energy is set at CaF + CaF asymptote.  We find that the asymptotes Ca$_2$ + F$_2$ and Ca$_2$F + F lie 73653 cm$^{-1}$ and 34483 cm$^{-1}$ above the incident CaF + CaF asymptote, and are therefore energetically inaccessible at low collision energies. 

On the other hand, for the dissociation channel CaF$_2$ + Ca, there are two possible asymptotes. The asymptote having  symmetry ($^3A^{''} + ^1S$) lies 10313 cm$^{-1}$ above from the CaF + CaF asymptote and is inaccessible. By contrast, the CaF$_2$ + Ca asymptote is lower in energy but possesses singlet symmetry, irrelevant to direct reaction on the triplet surface.  From these considerations, we conclude that the triplet surface is non-reactive.

\subsection{Four body surface}\label{glob}
 Construction of the potential energy surface of the four-atom CaF-CaF system is performed in two steps. Initially, an \textit{ab initio} calculation is carried out, as described above, on a  selected grid of atomic coordinates.  The surface at other configurations, where the {\it ab initio} calculation was not performed, is estimated using  Gaussian Process (GP) interpolation.  To this end, the coordinates of the atoms are cast in Jacobi coordinates ($R, r_{13}, r_{24}, \theta_1, \theta_2, \phi$) as shown in FIG.\ref{fig1}, where $R$ is the center of mass (CM) distance between two CaF molecules, $r_{13}$ ($r_{24}$) is the monomer bond length of CaF, $\theta_1$ ($\theta_2$)  is the polar angle, and $\phi$ is the dihedral angle.  
 
As a first step, we calculate global and local minima geometries of the triplet surface of the Ca$_2$F$_2$. This is accomplished by geometry optimization method using the relevant MOLPRO subroutine, and these calculations are done in the 
RHF-CCSD(T) level of theory with the same basis sets. We find one global minimum and one local minimum having symmetries $D_{2h}$ and $C_s$, respectively. The depth of the well ($D_e$) for the global minimum is 18074 cm$^{-1}$ whereas for the local minimum $D_e = 8760$ cm$^{-1}$, relative to the CaF + CaF asymptote. The geometrical parameters and the $D_e$'s are tabulated in Table.\ref{tb2}. 

\begin{table}[h!]
\centering
\caption{The energy of the PES Ca$_2$F$_2$ in cm$^{-1}$ for the global and local minima with optimized bond lengths (in Bohr). The energies are referred to the CaF + CaF dissociation threshold.}
\begin{tabular}{ |c|c|c|c|c| c|c|c|c|} 

\hline
surface &symmetry& $r_{12}$&  $r_{13}$ & $r_{34}$ & r$_{23}$  & r$_{24}$ &r$_{14}$ &E$_{min}$  \\
\hline
singlet &  D$_{2h}$& 6.360 & 4.057 & 5.039 &   -     &   -    &   -   & -18923 \\
triplet   &          & 6.402 & 4.065 & 5.013 &   -   &   -    &   -     &-18074 \\
\hline
singlet&  C$_s$   & 6.522 & 3.756 & 7.337 &  4.076   &  4.007 & 10.268  & -15363 \\
triplet  &        &7.035  & 3.765 & 7.144 &  4.120  &   3.949 & 10.628   &-8760 \\
\hline
\end{tabular}
\label{tb2}
\end{table}

In Table.\ref{tb2} we present the optimized bond lengths and optimized energies for the two minima of the singlet surface \cite{sardar2022} in addition to the triplet surface, to make a comparison. We note that both the states, singlet and triplet surfaces are minimized at nearly the same geometry and have essentially the same energy, with the singlet being only is 849 cm$^{-1}$ deeper. 


Qualitatively, the similarity in minima can be seen as follows.  Electrons migrate in the four-atom system so that the two F atoms are essentially closed-shell F$^-$ ions, while the positive Ca ions each have spin 1/2.  The singlet and triplet states of the pair of Ca ions would normally be expected to differ in energy, as the molecular orbital in the coordinate between them would be symmetric and bonding for the singlet state, antisymmetric and antibonding for the triplet state.  However, in the $D_{2h}$ equilibrium geometry, the Ca ions are a distance 6.4 a$_0$ apart, whereby both orbitals are near zero in the middle, hence minimizing the distinction between bonding and anitibonding orbitals.

\begin{figure}[htp]
    \centering
     \begin{subfigure}[b]{0.49\textwidth}
         \centering
         \includegraphics[width=\textwidth]{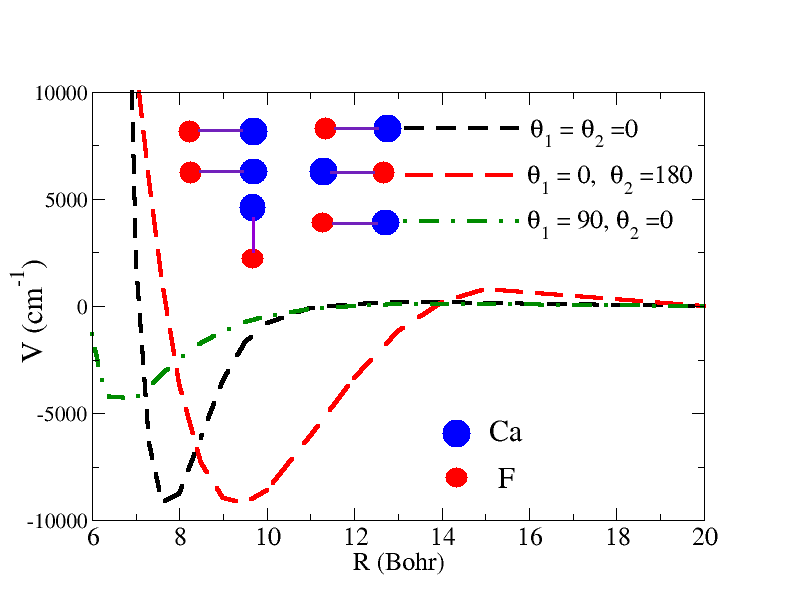}
         \label{fig}
     \end{subfigure}
     \hfill
     \begin{subfigure}[b]{0.49\textwidth}
         \centering
         \includegraphics[width=\textwidth]{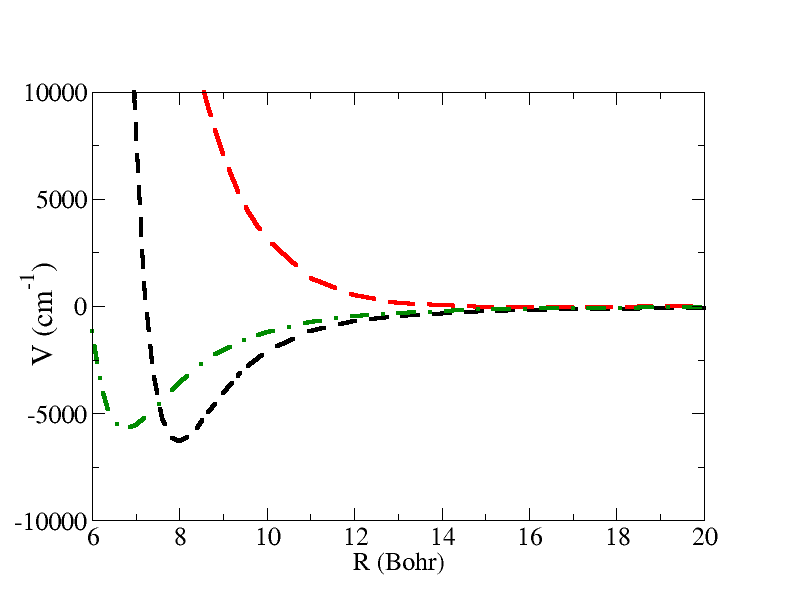}
         \label{fig:three sin x}
     \end{subfigure}
\caption{The variation of energies for the geometries head-tail, head-head, and T-shape are shown in the LHP and RHP for the singlet and triplet surface of CaF-CaF, respectively. Blue and red circles represent the atoms Ca and F. }     
\label{sing_trip}     
\end{figure}

Finally, we compare the singlet and triplet surfaces of CaF-CaF for some particular geometries, namely, head-tail ($\theta_1=0, \theta_2=0$), head-head ($\theta_1=0, \theta_2=\pi$), and T-shape ($\theta_1=\pi/2, \theta_2=0$) of the CaF-CaF. In the left-hand panel (LHP) of FIG.\ref{sing_trip}, we present a variation of energy of CaF-CaF for the singlet state as a function of $R$ (the distance between two centers of mass of CaF-CaF) for head-tail (black), head-head (red), and T- (green) geometries. The blue and red circles represent the atoms Ca and F, respectively.  Potential curves for the same geometries are  plotted in the right-hand  panel (RHP) for the nonreactive triplet surface. There is a noticeable difference between the head-head orientation between the singlet and triplet surface of CaF-CaF.   Notably, in the head-head geometry, the singlet potential is far deeper than the triplet potential.  This is as expected for the close approach of the essentially spin-1/2 Ca$^+$ ions, with the F$^-$ atoms behaving as spectators.

\section{Basis set extrapolation}\label{bsie}
Here we discuss the quality of our chosen basis set. To this purpose, we compare the global well depth obtained for our standard basis set for Ca (cc-pw-CVTZ-PP) to the other two correlation consistent basis sets having higher and lower angular momentum compared to the basis cc-pw-CVTZ-PP. The basis set of F = aug-cc-PVTZ is assumed to be the same for this calculation. We find that the correlation consistent basis having higher angular momentum of Ca (cc-pw-CVQZ-PP) offers the depth of the well  17992 cm$^{-1}$ for the  global minimum, in agreement  with our standard basis set for Ca as shown in Table.\ref{tb2}. However, the depth of the well becomes 5\% uncertain concerning another tested basis set of Ca = cc-pw-CVDZ-PP. Therefore, DZ basis set of Ca is not enough to calculate the complete surface.

Next, we make a quantitative estimation to verify further the quality of the used basis sets both for Ca and F in terms of the complete basis-set (CBS) extrapolation. The CBS extrapolation scheme is quite useful for large correlation- or polarization-consistent basis sets. The larger the basis sets we use, the more accurate the estimate of the CBS limit we will obtain. Basis set extrapolations are most important for highly correlated electronic structure methods, where basis set effects are large and often quite systematic in nature, and where calculations with large basis sets are often prohibitively expensive. 

Here we perform the extrapolation for the two basis sets, identified by whether they include triple (TZ, $n=3$) or else quadruple (QZ, $n=4$) excitations to explore the CBS limit. This basis set extrapolation provides an accurate estimate for the CBS limiting energy, which by definition has zero basis set superposition error (BSSE). Several formulas have been used for extrapolating finite basis set results to the complete basis set limit. An accurate and robust extrapolation functional is given by
\begin{equation}
    E_n = E_{CBS} + A(n+1/2)^{-4} 
    \label{eq1}
\end{equation}
where $A$ is the fitting coefficient and $n=3$ or $4$ as indicated above.  To estimate the CBS energy, we consider TZ basis both for Ca = cc-pw-CVTZ-PP and F = aug-cc-PVTZ, and we use the basis sets cc-pw-CVQZ-PP and aug-cc-PVQZ for Ca and F as QZ basis.  If we solve two equations obtained from Eq.\ref{eq1}, one for $n=3$ and one for $n=4$, and eliminating A, the CBS energy value becomes
\begin{equation}
    E_{CBS} = 1.5772 E_{QZ} - 0.5772 E_{TZ}
\end{equation}
Here, E$_{nZ}$ represents the energy corresponding with the basis sets nZ ($n = 3, 4$) for a particular geometry. We estimate the $E_{CBS}$ limiting energy -17974 cm$^{-1}$ for the global minimum of the triplet CaF-CaF surface. Therefore, the difference in energy between the depth of the well  for the TZ basis concerned with Ca and F and $E_{CBS}$ is approximately 100 cm$^{-1}$.  We conclude that the TZ basis set  is adequate for calculating the complete surface instead of the larger QZ acronym of the basis Ca and F.

\section{Gaussian process regression} 
Machine learning (ML) is a powerful tool  for interpolating multidimensional potential energy surfaces. One of the most familiar approaches to building interpolation of models is Gaussian Process regression, which is a non-parametric kernel-based probabilistic ML algorithm including Bayesian information criterion. 

For interpolation purposes, we convert atomic coordinates from their original Jacobi coordinates to inverse-atomic-distance coordinates.  Each set of six such coordinates is denoted by the six-dimensional vector ${\bf x_i}$.  Thus the total data set, for $N$ configurations at which the energy is computed, is denoted ${\bf x} = ({\bf x_1},{\bf x_2},....{\bf x_N} )^{T}$.  The $ab$ $initio$ energies are given by $\bf y$ for each ${\bf x_i}$.  The GP model is trained by ${\bf x}$ and the prediction of the energy as a normal distribution at an arbitrary point ${\bf x_*}$ is perceived by a mean ($\mu $) and standard deviation ($\sigma$) having the following form
\begin{equation}
 \mu({\bf x_*}) = K({\bf x_*},{\bf x})^T \left[K({\bf x},{\bf x}) + \sigma^2_N I \right]^{-1} \bf y 
\end{equation}
\begin{equation}
 \sigma({\bf x_*}) = K({\bf x_*},{\bf x_*}) - K({\bf x_*},{\bf x})^T\left[K({\bf x},{\bf x}+ \sigma^2_N I)\right]^{-1} K({\bf x_*},{\bf x})
\end{equation}
Here, the quantity $K({\bf x},{\bf x})$ is a square matrix with dimension $n \times n$ and the elements are $K(i, j) = ({\bf x}_i, {\bf x}_j)$. Therefore, the covariances between ${\bf y}({\bf x}_i)$ and ${\bf y}({\bf x}_j)$ are denoted by the matrix elements $ ({\bf x}_i, {\bf x}_j)$, where  $K$ is a kernel function that depends on some parameters. The parameters of the kernel functions are found by maximizing the log marginal likelihood (LML) which is given by 
\begin{equation}
 \log p({\bf y}|{\bf x\theta_3}) = -\frac{1}{2}{\bf y}^TK^{-1}{\bf y} - \frac{1}{2}\log|K| - \frac{N}{2}\log(2\pi)
\end{equation}
where $|K|$ is the determinant of K and $\theta_3$ is the collective set of parameters for the analytical function of the kernel. 

The quality of GP fit for a multidimensional surface depends on the kernel and the coordinate representation.  In our fitting method, we used a Mat$\acute{e}$rn kernel (M) \cite{rasmussen} having form
\begin{equation}
 M({\bf x}_i, {\bf x}_j) = \sum_{{k} = 1}^{6} \left[ 1 + \sqrt{5}l_k^{-1} |x_{i,k} - x_{j,k}| + \frac{5}{3}l_k^{-2}|x_{i,k} - x_{j,k}|^2 \right] \times  exp \left( -\sqrt{5}l_k^{-2} |x_{i,k} - x_{j,k}|^2 \right)
\end{equation}
Here, $l_1 -l_6$ are the characteristic length scales of Mat$\acute{e}$rn kernel, and are the parameters of $\theta_3$.

\subsection{Grids selection and symmetrization of GP}\label{sec:mysection}
In order to construct the training set for GP models, we construct
a grid in  Jacobi coordinates,  considered using Latin hypercube sampling (LHS) \cite{stein1987}. We consider a grid where both of the monomer bond lengths ($r_{13}$, $r_{24}$) of CaF  vary from 3.2 a$_0$ to 7.5 a$_0$ with $r_{24} > r_{13}$, the intermolecular separation $R$ varies from 4.0 a$_0$ to 20 a$_0$, and the angles $\theta_1$, $\theta_2$, and $\phi$ vary from 0 to $\pi$. 
We initially perform a RHF calculation on these random LHS grids.  We discard those points for which the HF energy lies above a certain threshold, these points being regarded as irrelevant to the dynamics at ultralow collision energies.    We use two criteria for the selection of the cutoff energy depending on the interparticle distances among the four atoms in Ca$_2$F$_2$ as described below:

1. When the CaF monomer bond length is smaller than 3.4 $a_0$ and the bond length between two Ca atoms is less than 10 a$_0$, we use the cutoff energy 4000 cm$^{-1}$ to account for the repulsive barrier appropriately.

2. Otherwise, we consider a lower value of cutoff energy, 1000 cm$^{-1}$.

We perform the full {\it ab inito} electronic structure calculation on the resulting grid, using the RHF-CCSD(T) method described above.   The \textit{ab initio} points on these grids are calculated only for $r_{13} < r_{24}$, while the remaining surface is constructed according to symmetry considerations.  

The procedure of symmetrization for the singlet potential of CaF-CaF is discussed in detail in our previous work \cite{sardar2022}. Here, we discuss the symmetrization criteria briefly. The symmetrization of the CaF-CaF surface arises either the exchange of two Ca nuclei, or exchange of the two F nuclei. This symmetrization procedure is accomplished by the exchange operators between two identical nuclei and a switching function.  We define two functions $ F_1 (\vec x)$ and $F_2 (\vec x)$ between which switch is done:
\begin{equation}
 F_m\left[u; m, w, F_1 (\vec x), F_2 (\vec x)\right] = y(u, c, w)F_1 (\vec x) + \left[1-z(u, c, w)\right]F_2 (\vec x) 
\end{equation}
where the sigmoid function $y$ should be twice differentiable and switches within the finite interval $(c-w) < u <(c+w)$, and it is given by \cite{arthur2019,sardar2022} 
\[
   y(u, c, w) =
  \Biggl\{ {\begin{array}{cc}
  \hspace{-5.2cm} 0  & \text{if} \hspace{0.2cm} u \le c-w   \\ 
   \frac{1}{2} + \frac{9}{16}{\sin\frac{\pi(u-c)}{2w} + \frac{1}{16}\sin\frac{3\pi(u-c)}{2w} }, & \text{if} \hspace{0.2cm} c-w < u < c+w  \\
   \hspace{-5.2cm} 1 & \text{if} \hspace{0.2cm} u \ge c+w \\
  \end{array} } 
\]
Here, $u$ is the parameterizing parameter for switching, $c$ is the value of $u$ around which switch is centered and $w$ is halfwidth of the switching interval. Finally, the symmetrization scheme for the CaF-CaF arrangement is given by
\begin{equation}
 V^{GP}_1(\vec x) = F_m\left[\frac{r_{13}}{r_{13}+r_{24}}; \frac{1}{2}, \frac{1}{16},  V^{GP}(\vec x),  V^{GP}(\hat P_{12}\hat P_{34}\vec x)\right]
\end{equation}
\begin{equation}
 V^{GP}_2(\vec x) = F_m\left[\frac{r_{23}}{r_{23}+r_{14}}; \frac{1}{2}, \frac{1}{16},  V^{GP}(\hat P_{12} \vec x),  V^{GP}(\hat P_{34}\vec x)\right]
\end{equation}
\begin{equation}
 V^{GP}_{CaF-CaF}(\vec x) = F_m\left[\frac{r_{13}+r_{24}}{r_{13}+r_{24}+r_{23}+r_{14}}; \frac{1}{2}, \frac{1}{16},  V^{GP}_1(\vec x),  V^{GP}_{2}(\vec x)\right]
\end{equation}
 where $p_{ij}$ are the exchange or permutation operators, which interchange the nuclei $i$ and $j$. This  symmetrization yields two advantages. First, symmetrization can reduce the configuration space over which we   perform the GP fit, and secondly, the addition of the symmetrically equivalent points to the training set can improve the quality of the GP fit around the symmetrization boundaries. 
 
\subsection{Construction of training set and interpolation} 

\begin{figure}[htp]
    \centering
     \begin{subfigure}[b]{0.54\textwidth}
         \centering
         \includegraphics[width=\textwidth]{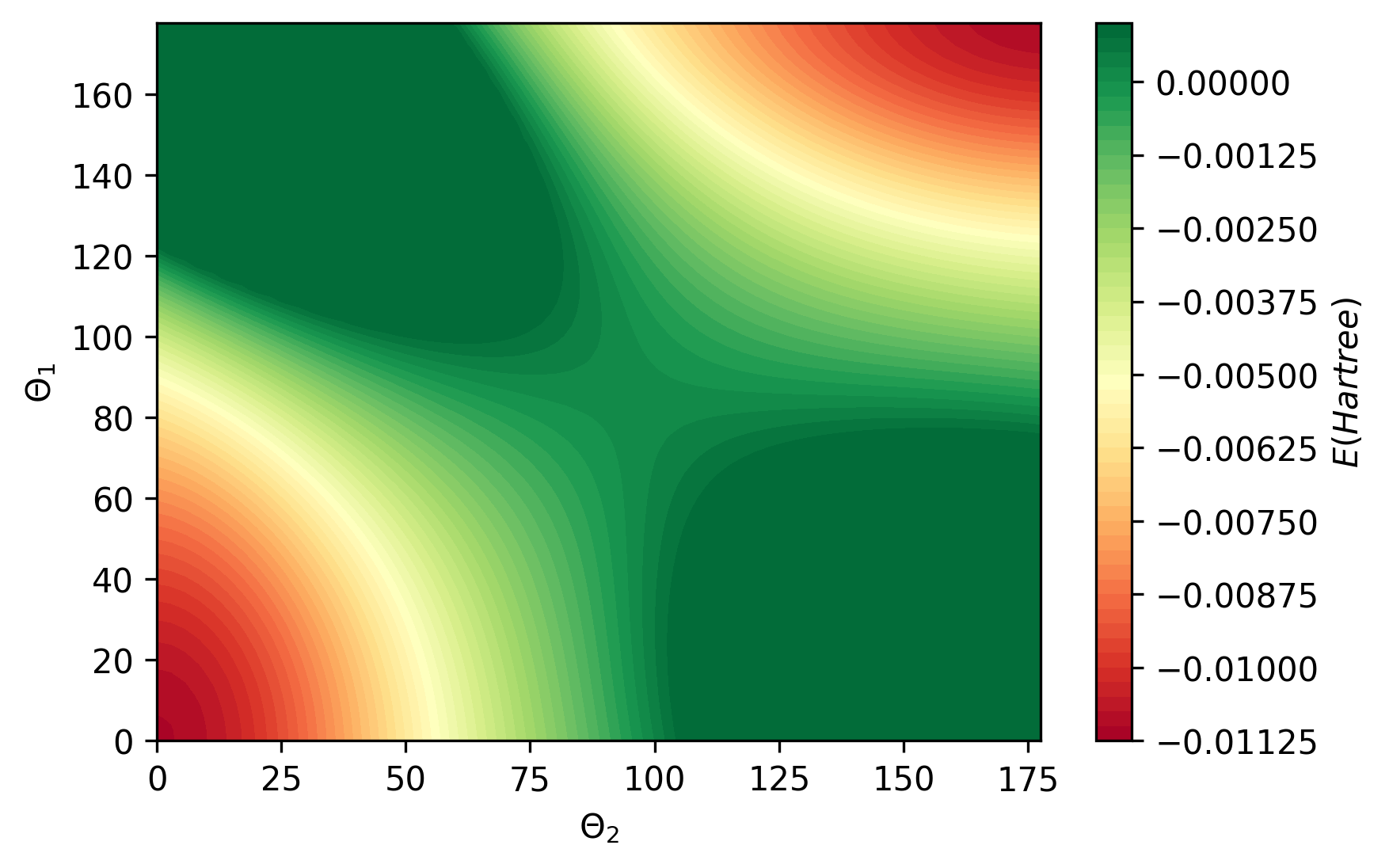}
         \label{fig}
     \end{subfigure}
     \hfill
     \begin{subfigure}[b]{0.45\textwidth}
         \centering
         \includegraphics[width=\textwidth]{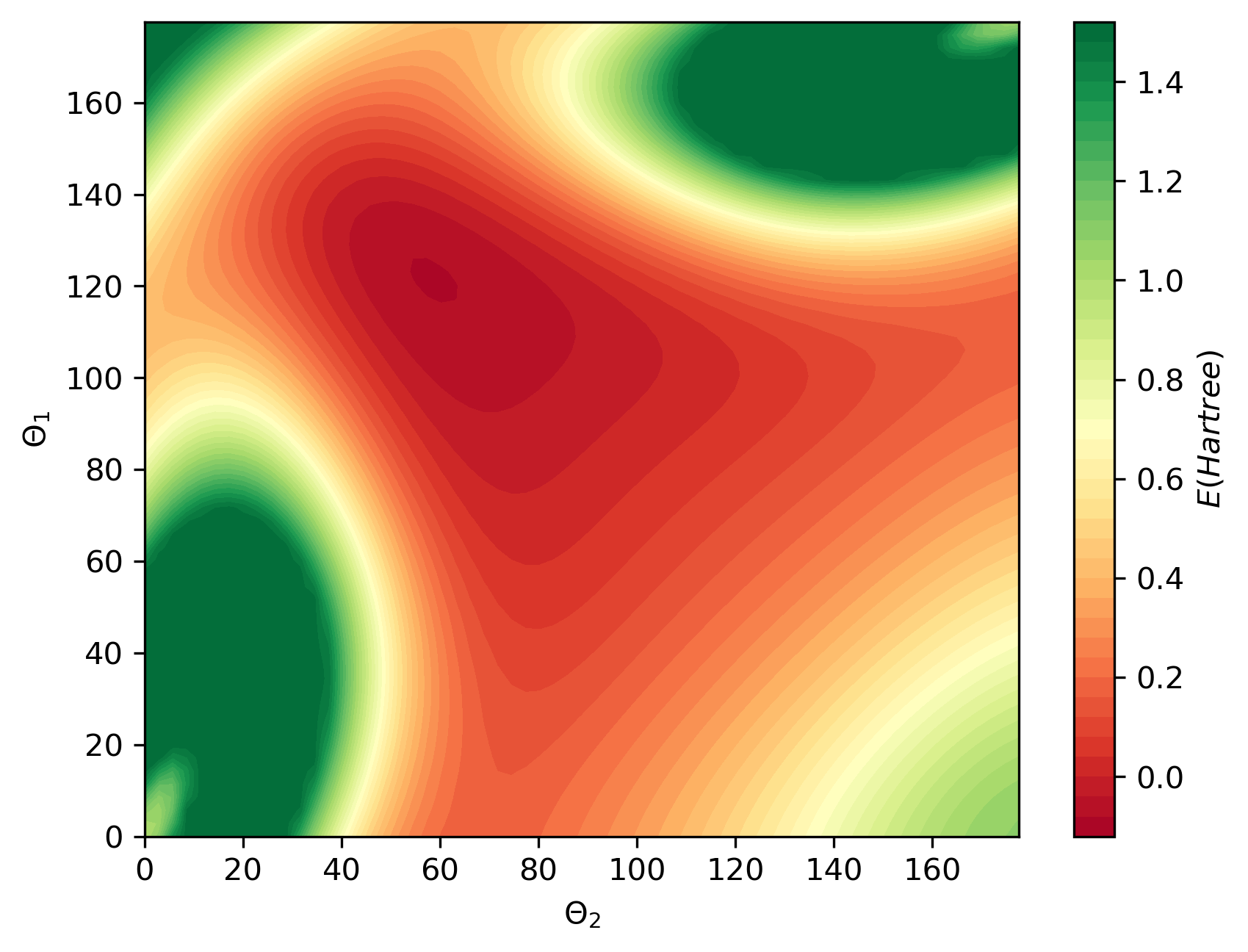}
         \label{fig:three sin x}
     \end{subfigure}
\caption{Variation of energy with respect to $\theta_1$ and $\theta_2$ in Jacobi coordinates, keeping other parameters remain constant. Left panel: $r_{13}=r_{14} = 3.7 a_0$, $R=10$, $\phi=180^{\circ}$.  Right panel: $r_{13}=r_{24} = 4.05 a_0$, $R=4.63 a_0$, $\phi = 180^{\circ}$. }     
\label{global}     
\end{figure}

Following the algorithm above, we constructed a training data set based on the \textit{ab initio} calculation for a total of 2075 atomic configurations. The coordinates of the grid points are transformed to the inverse atomic distance coordinates. These coordinates are convenient, since
it is difficult to describe different chemical arrangements of the four-body system equivalently by Jacobi coordinates making it is troublesome to fit the complete PES \cite{sardar2022}. On the other hand, the GP fit works well everywhere on the surface exploiting inverse atomic distance coordinates. 

Thereafter, we add symmetrically equivalent points to this \textit{ab initio} data set, making a total number of grid points 3136 in the data set. We divide these points randomly into training and test sets, comprising  80\% and 20\% of the data, respectively.  We fit these training set of data by GP regression. For this fitting, the kernel parameters of the Mat$\acute{e}$rn kernel become [0.705, 0.74, 0.751, 0.749, 0.741, 1.08], the log marginal likelihood equal to 11190. The quality of the fit can be analyzed by computing the root-mean-square error of the interpolated data, with respect to the test data set. This error of the  fitting is on the order of 300 cm$^{-1}$, setting an approximate uncertainty of the fit. The interpolated potential appears on the  \href{https://github.com/jila2021/Triplet-CaF-CaF.git}{Github site}.

As a second check, we find the minima of the interpolated surface, using a steepest-gradient-descent optimization method. The optimized results in terms of the energy and bond lengths  are listed in Table.\ref{tb-opt}. The optimized energy for these minima specifically for the global minimum is in better agreement with the optimized results in MOLPRO with RHF-CCSD(T) calculation with an uncertainty approximately 100 cm$^{-1}$. This results indicate that the number of training points is sufficient for the construction of the triplet surface CaF-CaF. 

As a further check we present a contour diagram in FIG.\ref{global} in Jacobi coordinates to see the variation of energies for some special constricted geometries and detecting the global minimum of the Ca$_2$F$_2$ surface using the GP model. The purpose of these contour plots is to explore whether our GP model can reproduce the known results of the triplet surface of CaF-CaF obtained from MOLPRO. 
In the left hand panel of FIG.\ref{global}, we fix two monomer bond lengths ($r_{13}$ and $r_{24}$) of each CaF at their equilibrium internuclear distance, two CaF molecules are separated by a distance $R=10$ Bohr from the center of mass. Then we vary polar angles ($\theta_1$ and $\theta_2$) from 0-$180^0$ keeping dihedral angle $\phi=0$. The potential is minimum at the head-to-tail geometry, as expected.

In the right hand panel of FIG.\ref{global}, we fix $r_{13} = r_{24} = 4.05$ Bohr, $R = 4.63$ Bohr, $\phi = 180^0$, and we vary $\theta_1$ and $\theta_2$ from $0-180^0$.  The minimum energy shown is at the geometry $\theta_1 = 121^0$ and $\theta_2 = 59^0$, nearly coinciding with the result of the MOLPRO minimization.  

\begin{table}[h!]
\centering
\caption{The optimized energy (in cm$^{-1}$) and bond lengths (in Bohr) of the interpolated PES Ca$_2$F$_2$  for the global and local minima. }
\begin{tabular}{ |c|c|c|c|c| c|c|c|c|} 

\hline
symmetry & $r_{12}$&  $r_{13}$ & $r_{34}$ & r$_{23}$  & r$_{24}$ &r$_{14}$ &E$_{min}$  \\
\hline
 D$_{2h}$ & 6.360 & 4.050 & 5.016 &   -     &   -    &   -   & -18137 \\
 C$_s$    & 7.265 & 3.752 & 6.615 &  4.147   &  3.916 &  10.380  & -8477 \\
\hline
\end{tabular}
\label{tb-opt}
\end{table}
\section{Lifetime of the collision complex}
A recent surprise in the field of ultracold molecular collisions is that the lifetime of the collision complex can be quite large.  The simplest understanding of this is as follows.  The four-body complex comprises a large phase space volume or equivalently, a high density of states $\rho$.  If the number of open channels available for the complex to decay into is denoted $N_o$,  then a ready approximation for the lifetime of the complex is given by the Rice-Ramsperger-Kassel-Marcus (RRKM) expression 
\begin{equation}
    \tau = \frac{h\rho}{N_o}.
\end{equation}  
In the case of ultracold molecules prepared initially in their ground state (an achievable situation these days), the number of open channels is only $N_o=1$, often leading to estimated RRKM lifetimes of microseconds to milliseconds.  In the case where lifetimes have been measured, they are sometimes even far larger, for reasons that remain mysterious.  The current situation is reviewed in Bause, {\it et al} \cite{Bause23_JPCA}. 

Given a potential energy surface such as our triplet potential, the density of states can be estimated by a quasiclassical calculation derived in Ref. \cite{PhysRevA.100.032708}. For a system having $N_i$  particles, the total number of quantum states below a certain energy $E$, with total angular momentum $J_0$ and center of mass ${\bf X} = (0, 0, 0)$ is given by
\begin{equation}
    N(E, J_0)=\frac{1}{h^{3N-3}\Pi_i N_i!}\int d{\bf x}\int d{\bf p} {\hspace{0.1cm}} \theta[E-H({\bf x}-{\bf p})]\times \delta[{\bf P}({\bf p})]{\hspace{0.1cm}} \delta[{\bf X}({\bf x})]{\hspace{0.1cm}} \delta[{\bf J_0}-{\bf J(x,p)}]
\end{equation}
Here, $\theta(E)$ is the Heaviside step function, with $\theta(x)=1$ for $x\ge 0$ and $\theta(x)=0$ for $x< 0$. The factor $\Pi_i N_i!$ is used to make correction for indistinguishability of the particles. The derivative of $N(E, J_0)$ with regard to energy E, gives the DOS. 
\begin{figure}[htp]
    \centering
    \includegraphics[width=9cm]{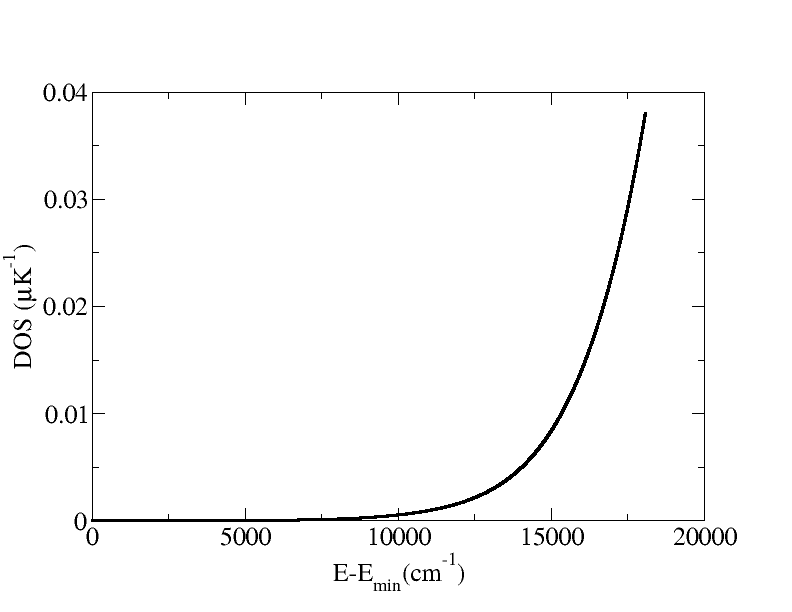}
    \caption{Density of state as a function of energy for the triplet CaF-CaF. }
    \label{dos}
\end{figure}
For the four atom system, triplet CaF-CaF we use Jacobi coordinates which can be expressed as ${\bf q} = (r_{13}, r_{24}, R, \theta_1, \theta_2, \phi)$. The DOS in these coordinates can be expressed as \cite{PhysRevA.100.032708}
\begin{equation}
    \rho(E) = \frac{g_N4\pi^6(2J+1)m_{Ca}^6m_{F}^6}{h^9(2m_{Ca}+2m_F)^3g_{Ca_2F_2}}\int\frac{r_{13}^4r_{24}^4R^4\sin^2(\theta_1)\sin^2(\theta_2)}{\det\cal I({\bf q})\sqrt{\det\cal A({\bf q})}}[E-V({\bf q})]^2 d{\bf q}
\end{equation}
The terms $g_{Ca_2F_2}$ is associated with degeneracy to account for indistinguishability, and $g_N$ is a quantum mechanical factor which is defined as the fraction of classical phase space associated with parity. The masses of of the atoms Ca and F are assigned by $m_{Ca}$ and $m_F$, respectively. The explicit expressions for the terms $\cal I({\bf q})$ and $\cal A({\bf q})$ are given in reference \cite{PhysRevA.100.032708}.

In the quasiclassical calculation of the DOS for the triplet potential energy surface of CaF-CaF, we choose an integration grid ranging from $r_{13}$ ($r_{24}$) from 4$a_0$ to 10$a_0$ with a spacing of 0.3$a_0$, and for $R$ we use a grid of 71 points placed from 4$a_0$ to 18$a_0$. We use a 20-point Gauss-Legendre quadrature for $\theta_1$ and $\theta_2$, and a 8-point Gauss-Chebyshev quadrature is used for $\phi$ between 0 to $\pi$. Note that, we use $r_{24}>r_{13}$ and multiply the result by a factor of 2 due to symmetry, and an additional 2 factor should be included for compensating $\phi$ for running up to $\pi$ instead of $2\pi$ \cite{PhysRevA.100.032708}. Next, a geometry-dependent weighting factor needs to be assigned to the integrand, and this weighting factor $W({\bf q})$ is deduced based on the symmetrization of the surface. 
For the triplet state arrangement of CaF-CaF, $W({\bf q})=W_1 W_2$, where the weighting factor $W({\bf q})$ is equivalent to the function $y(u, c, w)$  defined in the previous section. Here, $W_1 = W(u_1, 1, 1/4)$ with $u_1=\frac{r_{12}+r_{34}}{2(r_{13}+r_{24})} + \frac{r_{12}+r_{34}}{2(r_{23}+r_{14})}$, and $W_2 = W(u_2, 1/2, 1/16)$ with $u_2=\frac{r_{23}+r_{14}}{r_{13}+r_{24}+r_{23}+r_{14}}$.

Finally in Fig.\ref{dos}, we present the DOS as function energy $E-E_{min}$ in cm$^{-1}$, where $E_{min}$ is energy of the minimum of the potential. The DOS is found to be 0.038 $\mu$K$^{-1}$ near the dissociation energy for $J=0$ in the field free case, and the corresponding RRKM sticking time of the collision complex is 18 $\mu$s. The original code of the DOS is written by Arthur \textit{et al} \cite{PhysRevA.100.032708} and is availabe on the github webside \href{https://gitlab.science.ru.nl/theochem}{Github site}.

This lifetime is comparable to those associated with midsize alkali dimer complexes, such as KRb.  The effect of sticking in these collisions is to temporarily remove molecules from an ultracold gas, as if they are reacting chemically, even when they are not.  If the collisions take place in an optical dipole trap, the complex may absorb a photon of the trapping light and be truly removed from the trap.  

Even if this is not the case, a molecule such as CaF might be in danger.  If the molecules remain in the triplet potential, they may survive the complex and re-emerge as CaF molecules through the single open channel.  However, if the molecules become stuck long enough in a collision complex in the triplet state, they will have multiple opportunities during  the lifetime of the complex to escape to the singlet surface by spin-orbit interactions, and hence complete the chemical reaction.

\section{Conclusion}
We constructed an  \textit{ab initio} potential energy surface of triplet CaF-CaF,  followed by gaussian process interpolation. The \textit{ab initio} method used the  CCSD(T)/cc-pwCVTZ-PP and aug-cc-PVTZ level of theory. The deduced potential was used to determine the DOS via quasiclassical calculation, and subsequently used to determine the RRKM sticking time of 18 $\mu$s of the collision complex. The knowledge of the DOS, and RRKM lifetime would be useful to understand the loss mechanism of the complex.

\begin{acknowledgement}
  We also acknowledge funding from an AFOSR-MURI grant as grant number GG016303.
Dibyendu Sardar especially acknowledges discussions with Kirk A. Peterson and Tijs Karman for \textit{ab initio} calculations, and thankful to Arthur Christianen for initial discussions about the gaussian process regression.
\end{acknowledgement}

\bibliography{ca2f2}

\end{document}